\documentclass[prl,twocolumn,amsmath,nobibnotes,nofootinbib,superscriptaddress]{revtex4}

\usepackage{bm} \usepackage{graphicx} \usepackage{epstopdf}
 \def\ba{\begin{eqnarray}} \def\ea{\end{eqnarray}}
\def\be{\begin{equation}} \def\ee{\end{equation}} \def\({\left(}
\def\){\right)} \def\[{\left[} \def\]{\right]} \def\<{\left<}
\def\>{\right>}

\def\ba{\begin{eqnarray}}
\def\ea{\end{eqnarray}}
\def\be{\begin{equation}}
\def\ee{\end{equation}}
\def\({\left(}
\def\){\right)}
\def\[{\left[}
\def\]{\right]}
\def\<{\left<}
\def\>{\right>}

\newcommand{\rar}{\rightarrow}

\begin{document}

\title{Measures on transitions for cosmology from eternal inflation }
\date{\today}

\author{Anthony Aguirre}
\email{aguirre@scipp.ucsc.edu}
\affiliation{SCIPP, University of California, Santa Cruz, CA 95064, USA}
\author{Steven Gratton}
\email{stg20@cam.ac.uk}
\affiliation{Institute of Astronomy, Madingley
 Road, Cambridge, CB3 0HA, UK}
\author{Matthew C Johnson}
\email{mjohnson@physics.ucsc.edu}
\affiliation{SCIPP, University of California, Santa Cruz, CA 95064, USA}

\begin{abstract}
We argue that in the context of eternal inflation in the
landscape, making predictions for cosmological -- and possibly particle physics -- observables requires a measure on the possible cosmological histories as opposed to one on the vacua themselves. If significant slow-roll inflation occurs, the observables are generally determined by the history after the last {\em transition} between metastable vacua.  Hence we start from several existing measures for counting vacua and develop measures for counting the transitions between vacua.   
\end{abstract}

\maketitle

\section{Introduction}
\label{sec-intro}

"Cosmic inflation", the idea that the early universe underwent an epoch of accelerated expansion, was developed to account for the universe's observed uniformity, geometric near-flatness, absence of GUT monopoles, and required small density inhomogeneitites.  But while inflation grants these wishes, it, like the proverbial genie let out of the bottle, is difficult to contain. In nearly any model in which the scalar field potential driving inflation has multiple minima, the very exponential expansion responsible for inflation's predictive successes also prevents a global end to inflation: the expansion shields still-inflating regions  from the encroaching effect of those where inflation has ended.  Such models can be fairly described as "eternal" because a time foliation exists in which the physical inflating volume expands exponentially forever, and inflation only ends {\em locally} in regions where the field settles into a particular, low energy, potential minimum or ``vacuum".

Moreover, developing understanding of metastable states in string theory seems to be pointing towards a vast, interconnected, many-dimensional web or ``landscape'' of many, many such vacua. Populated by eternal inflation, this would lead to an ensemble of ``bubble universes" with diverse properties, making predictions of low-energy observables probabilistic. A major open question is how -- even in principle -- this probability distribution should be calculated, and significant effort has been expended in finding methods to assign probabilities $P(v_k)$ to different vacua $v_k$ using, e.g., bubble abundances, frequencies of vacuum entries, and probability currents (e.g.,~\cite{Winitzki:2005ya,Vilenkin:2006xv,Garriga:2001ri,Bousso:2006ev,Vanchurin:2006qp, Garriga:2005av,Easther:2005wi,Linde:2006nw}).

We argue here that such $P(v_k)$ are insufficient: while many
particle-physics-type observables may depend on the vacuum alone, many
cosmological observables depend not just on what vacuum a region is
in at some time, but also on {\em the history of that region}. Thus, what is actually required in principle is a measure over {\em histories} rather than over
vacua. Putting measures over histories is not a new concept
(e.g.,~\cite{Gratton:2005bi,Gibbons:2006pa}), but counting full histories to determine low-energy observables is probably overkill if significant inflation occurs after most transitions that lead to low energy vacua. The final transition type will typically determine the slow-roll inflationary history down to a low energy state, and hence answer most cosmological and low-energy particle physics questions. Thus a measure over {\em transitions} should be sufficient (and much simpler to calculate) for most purposes. 

In the remainder of this paper we support
this argument with specific examples and then, starting from two recent
proposals for calculating $P(v_k)$, develop measures for
probabilities of transitions. Note that we focus here
on eternal inflation as driven by multiple minima in a
scalar potential; it would be interesting and important to
extend the measures under discussion to treat Òstochas-
ticÓ eternal inflation (e.g., ~\cite{Linde:1993xx}) in the landscape as well.

\section{Transitions rather than vacua}
\label{sec-transitions}

In the ``multiverse" picture suggested by eternal inflation, the
20-odd parameters $\alpha_i$ defining both a ``standard model'' of particle
physics and a cosmology since inflation's end (see, e.g.~\cite{Tegmark:2005dy} for a listing) might be described by a 20-odd dimensional joint probability
distribution ${\cal P}_X(\alpha_i)$, where $X$ is some
``conditionalization object" such as a point in space, baryon, galaxy,
or ``observer", and ${\cal P}_X(\alpha_i)$ governs the chance -- given
no other information -- that an ``X'' inhabits a region with
parameters described by
$\alpha_i$~\cite{Aguirre:2005cj,Aguirre:2004qb}.

How can ${\cal P}_X(\alpha_i)$ be calculated?  A method based on vacua, as is generally done, might run as follows.  Suppose there is a unique
set $\alpha_i(w_k)$ of parameter values and a fixed number
$N_X(w_k)$ of X-objects associated with each $w_k$, where each $w_k$ is equated with a particular vacuum $v_k$. Then for each $i$ we might
calculate:
\be
{\cal P}_X(\alpha_i)=\int d\alpha_i' \sum_{k} N_X(w_k) P(w_k)\delta (\alpha_i'-\alpha_i(w_k)),
\label{eq-px}
\ee
normalize, and smooth the distribution if desired.  

\begin{figure}
\includegraphics[scale=0.42]{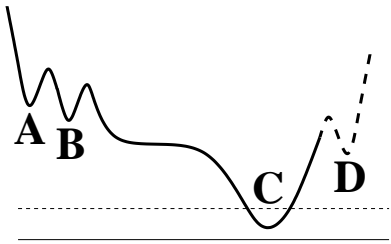}
\caption{
  \label{potential}
A simple potential landscape. We consider both a positive and negative
energy $C$-well, with the zeros in energy density denoted by the solid
or dotted line. In the text, we discuss both three-well (composed of
$(A,B,C)$) and four-well (composed of $(A,B,C,D)$) landscapes.}
\end{figure}

However, both $\alpha_i$ and $N_X$ often depend not just on the vacuum $v_k$, but also on how that vacuum was reached -- that is, there is a one-to-many mapping from vacua $v_k$ to observables $\alpha_i$. For example, consider the
potential $V(\phi)$ in Fig.~\ref{potential}. Bubbles of vacuum
$C$ can form via Coleman-De Luccia transitions~\cite{Coleman:1980aw}
from either the $B$ or $D$ vacua. The endpoint of tunneling from $B$
would lie on the flat region of the potential, whereas the endpoint of
tunneling from $D$ might be very near $C$'s minimum. The number $N_e$
of inflationary e-folds between tunneling and reheating then depends
on which of these two transitions took place.

Clearly, each vacuum will not correspond to a unique set of
$\alpha_i$ for inflationary predictions like $N_e$, the tensor/scalar
ratio, the curvature scale, the
perturbation amplitude, the reheating temperature, etc.\footnote{This
dependence can be seen ``in action" in Tegmark's
study~\cite{Tegmark:2004qd} of inflationary predictions for random
one-dimensional potentials.}  Instead, each vacuum maps to a set of
possibilities that may be large, given that in a many-dimensional
landscape there might be hundreds of directions from which to tunnel.
It is possible that some non-inflationary predictions could depend on
the vacuum tunneled from as well.
 
Even if a parameter $\alpha_{(i)}$ does depend on just the vacuum
$v_k$ (for instance the late-time vacuum energy), the ``counting
factor" $N_X$ very likely will not. For example, $X$ choices of ``a
unit volume on the reheating surface" (e.g.,~\cite{Garriga:2001ri}) or
``a galaxy" (e.g.,~\cite{Garriga:2002tq,Tegmark:2005dy} ) or ``a unit
of entropy generation"~\cite{Bousso:2006ev} would all seem to depend
on (at least) $N_e$.\footnote{In fact, $N_X$ will often be infinite
and require regularization; we will not pursue that thorny difficulty
here.} Thus even if $\alpha_{(i)}$ is merely {\em correlated} in
$N_X$ or $P$ with an observable that depends on the predecessor
vacuum, properly predicting $\alpha_{(i)}$ using ${\cal P}_X$ 
requires accounting for the transition history.

These considerations suggest that we may still use Eq.~\ref{eq-px} to calculate ${\cal P}_X(\alpha_i)$, but that each $w_k$ should correspond to a {\em transition} between two vacua, which will (in most cases) map directly both to a unique set of observables $\alpha_i$ and to a unique counting factor $N_X$. We stress here that fully labeling a transition requires specification of both a ``before" and an ``after" vacuum.

It is also quite possible that the transition rates and mechanisms
themselves depend on the transition history. For example,
Tye~\cite{Tye:2006tg} has recently argued that in a landscape like Fig.~\ref{potential}, very fast ``resonant" tunneling from $A\rightarrow C$ can occur if 
(a) the (non-resonant) $B\rightarrow C$ transition rate approximately equals the $A\rightarrow B$ rate, and (b) the shape of the potential near vacuum $B$ satisfies a ``resonance condition"  (see also~\cite{Baacke:2005gq}).  Such tunneling in a general landscape can be accounted for consistently, but only if one allows the $B\rightarrow C$ transition rate to depend upon whether or not the previous transition was $A\rightarrow B$. 

As another example, consider the case where the $C$-well in
Fig.~\ref{potential} has negative energy. A transition $D\rightarrow
C$ might (if the potential is suitably chosen) yield no post-tunneling inflation and lead quickly to a big-crunch in bubble's interior, so that the tunneling rate out of $C$
would either be extremely suppressed or even vanish identically.
But after a tunneling from $B \rightarrow C$, transitions back to $B$ might occur during the near-de Sitter phase during slow roll toward $C$.\footnote{This distinction would be critical in measures that yield very different probabilities depending on whether transitions are allowed -- with whatever probability -- out of a vacuum or not; see~\cite{Aguirre:2006ak}.}

\section{Transitions on a single worldline}

We now develop a transition-based analogue of Bousso's ``holographic probability'' measure for vacua~\cite{Bousso:2006ev}.\footnote{Indeed, the dependence of entropy production on the parent vacuum, and the necessity to introduce a formalism such as that presented here, was anticipated in the conclusions of the pre-print versions of~\cite{Bousso:2006ev}.}
Consider a worldline that passes through spacetime regions
described by different vacua. 
If we denote by $NM$
the transition from vacuum $M$ to $N$, then we can denote by
$p_i^{\alpha}$, with $\alpha=NM$, the probability that the $i$th
transition experienced by the worldline is $NM$. 

If we now assume that the probability that a transition $\beta$ is
followed by a transition $\alpha$ is independent of transitions before $\beta$, and 
denote this probability (or ``branching ratio") by $\mu_{\ \beta}^{\alpha}$, then the
$p_i^{\alpha}$ form a 
Markov chain, with   
\be
p_{i+1}^{\alpha}=\sum_\beta \mu^{\alpha}_{\ \beta}\,p_i^{\beta}.
\ee
Note that if $\alpha=NM$ and $\beta=LK$ then $\mu^{\alpha}_{\ \beta}$ is nonzero only for $L=M$, and that in general $\mu^{\alpha}_{\ \beta}\neq \mu^{\beta}_{\ \alpha}$. Also, normalization of the probabilities requires that $\sum_\alpha \mu^{\alpha}_{\ \beta}=1$
if transition $\beta$ ends in a metastable vacuum; if $\beta$ ends instead in a ``terminal" vacuum (which cannot be transitioned out of), $\mu^{\alpha}_{\ \beta}$ vanishes for all $\alpha$.

Now, if we start with some initial condition $p_0^{\alpha}$, and write ${\bm p}$ as a vector and ${\bm \mu}$ as a matrix (with entries labeled by the greek indices), then
${\bm p}_i=({\bm \mu})^i{\bm p}_0$, and the expected number $n^\alpha_j$ of transitions of type $\alpha$ after $j$ steps (excluding the ``zeroth'' transition) is: 

\be
\bm{n}_j = \sum_{i=1}^j \bm{p}_i = \bm{S}_j \bm{p}_0,
\ee
where $\bm{S}_j \equiv \sum_{i=1}^j (\bm{\mu})^i$.

The sum can be performed exactly if the landscape is terminal, and
must be regulated in the case of a fully recycling landscape (we refer
the reader to the Appendix of~\cite{Aguirre:2006ak} for analogous
details). In either case, in the ${j\rar\infty}$ limit the number of transitions is proportional to
\be
\bm{n}_{\infty} \propto \left\{ \text{adj} (\bm{1}-\bm{\mu}) \right\} \bm{\mu} \bm{p}_0,
\ee
where $\text{adj}$ denotes the adjoint matrix operation (i.e.\ the transpose of the matrix of cofactors of the matrix in question). Normalizing $\bm{n}_{\infty}$ yields probabilities for the various transitions in the model.  

To illustrate this method, consider a landscape with three vacua $(A,B,C)$, with vacuum energies $V_C < V_A, V_B$ (solid curve in Fig.~\ref{potential}),  that can experience nearest-neighbor transitions only. If $p_i^\alpha=(p_i^{AB},p_i^{BA},p_i^{CB},p_i^{BC})$, we obtain
\be
\bm{\mu}=
\begin{pmatrix}
0 & \mu_{\  BA}^{AB} & 0 & \mu_{\  BC}^{AB} \\
\mu_{\  AB}^{BA} &  0 & 0 & 0 \\
0 & \mu_{\  BA}^{CB} & 0 & \mu_{\  BC}^{CB} \\
0 & 0 & \mu_{\  CB}^{BC} & 0 
\end{pmatrix}.
\ee
Imposing the normalization condition on the columns, we obtain $\mu_{\ AB}^{BA}=1$,  $\mu_{\ BA}^{AB}=\epsilon$, $ \mu_{\ BA}^{CB} = 1-\epsilon$, $\mu_{\ BC}^{AB} = \delta$, $ \mu_{\ BC}^{CB}=1-\delta$, and $\mu_{\ CB}^{BC}=0$ (resp. $\mu_{\ CB}^{BC}=1$) if C is terminal (resp. recycling), with free parameters $\epsilon,\delta < 1$.

As an example, if the fictitious zeroth transition is $BA$ (i.e.\
$p^\alpha_0=(0,1,0,0)$), starting us in vacuum $B$, the expected number of transitions for the terminal case ($\mu_{\ CB}^{BC}=0$) is
$
\bm{n}_\infty = (\epsilon/(1-\epsilon) , \epsilon/(1-\epsilon) ,1, 0).
$
Normalizing, the transition probabilities are given by
\ba
&&P(AB) = \frac{\epsilon}{1+\epsilon}, \ \  P(BA)=\frac{\epsilon}{1+\epsilon}, \nonumber \\ 
&&P(CB)=\frac{1-\epsilon}{1+\epsilon}, \ \ P(BC)=0. 
\ea

Using any initial condition, we can compute the number of transitions in the recycling case ($\mu_{\ CB}^{BC}=1$), yielding $\bm{n}_\infty \propto (\delta,\delta,1-\epsilon,1-\epsilon)$.
Normalizing, the probabilities assigned to the various transitions are then:
\ba
&&P(AB) = P(BA)=\frac{\delta}{2(1-\epsilon + \delta)},\nonumber \\ 
&&P(CB)=P(BC)=\frac{(1-\epsilon)}{2(1-\epsilon + \delta)} .
\ea
 
\subsection{Recovery of one-point statistics}
 
Let us quantify the extent to which the transition counting measure presented above is a generalization of Bousso's~\cite{Bousso:2006ev} measure for vacua. In~\cite{Bousso:2006ev}, one considers $n_{v}$ vacua with transitions between them, the rates of which depend only upon the
starting and ending vacuum.  Describing these transitions requires
$n_v(n_v-1)$ transition rates with $n_v$ normalization conditions,
hence $n_v(n_v-2)$ independent numbers must be specified. In contrast,
there are $n_v(n_v-1)^2$ possible transitions between transitions,
with $n_v(n_v-1)$ normalization conditions, hence $n_v(n_v-1)(n_v-2)$
independent parameters. There is thus $n_v-1$ times as much freedom,
essentially corresponding to the $n_v-1$ ways a vacuum might be
entered.

Now let us see how probabilities for states can be
reproduced. Thinking in terms of states rather than transitions
suggests two things: (1) assuming that transition rates depend only
upon the initial and final states (that is, for a given $\alpha$, $\mu_{\
  \beta}^\alpha$ is identical for all $\beta$ that end in the same
state) and (2) that we are interested primarily in the probability
accorded to each vacuum $M$. To obtain this probability, we
simply sum $\bm{n}$ over all transitions that end in $M$, do likewise for all other vacua, then normalize. 

Probabilities for the states $A$, $B$ and $C$ in the examples above
can be found by setting $\delta = \epsilon$ (assumption (1) above) and
summing over the two transitions that end in $B$ to obtain 
results in agreement with those of~\cite{Bousso:2006ev}.

It is worth noting that under assumption (1), one can calculate the
relative frequencies $p(NM)$ of the different transitions by first
calculating the relative frequencies $p(M)$ of different parent vacua (but now including the starting transition in $\bm{n}_j$), then multiplying by the ``branching ratio'' $\mu^N_{\ M}$, which is the (normalized) probability that $M$ transitions into $N$. In cases where
assumption (1) holds, this can provide a simpler procedure.

\subsection{Higher moments and longer histories}
In principle it is possible that either (a) we might desire
probabilities for strings of three or more transitions, or (b)
transition rates might depend on the last two or more transitions. Probabilities for long chains are simple if transition rates depend
only on at most the previous transition. Then, if we wish to count
chains $PON...MLK$ along a worldline, we simply multiply $p(LK)$ by a
string of branching ratios:  $p(PON...MLK)= \mu_{\ ON}^{PO}...\mu^{ML}_{\ LK}p(LK)
$. 

If transition rates {\em do} depend on two or more previous
transitions, it is still straightforward to generalize the counting to longer histories (groups of transitions). Focusing on the count along a single worldline, if we
set $\alpha=PO...LK$, $\beta=NM...JI$, then $\mu_{\ \beta}^{\alpha}$
implements transitions from the transition group $I\rar ... \rar N$
into the group $K \rar ... \rar P$. This allows the transition rate to a new vacuum to depend on a history of transitions of arbitrary length. To accomplish this, we set $\mu_{\ \beta}^{\alpha}=0$ unless $\alpha=QNM...J$ for some $Q$; that is, we only allow transitions such as $CBA \rightarrow DCB$ or $DCB \rightarrow EDC$ but not, e.g. $CBA \rightarrow EDC$ (which would allow the $DC$ transition rate to depend on what transition occurs {\em after} $DC$). With this setup, we can calculate $p(N...M)$, using the same Markov chain techniques described above.

\section{Counting total transition numbers}
\label{sec-chc}
The measure discussed in the previous sections assigns weight to
various transitions occurring on a single worldline. It is also possible
to define a measure based on the total number of transitions occurring
in the eternally inflating spacetime. Consider the method of Garriga et al.~\cite{Garriga:2005av}, which
follows the evolution of a congruence of hypersurface-orthogonal
geodesics extending from some initial spacelike slice. The formalism
first calculates the fraction of geodesics in a given phase as a function of time. To extend this method, we must keep track of the fraction $f^{NM}$ of these ``comoving observers" in vacuum $N$ that came from vacuum $M$, such that 
$
\sum_{N,M} f^{NM} = 1.
$
The dynamics are determined by the rate equations
\be
\frac{df^{NM}}{dt} = - \left( \sum_{P} \kappa_{\ NM}^{PN} f^{NM} \right) + \left(\sum_{L} \kappa_{\ ML}^{NM} f^{ML} \right),
\label{eq-rates}
\ee
where $\kappa_{\ BA}^{CB}$ are the transition rates. 

The state-based rate-equation formalism can be recovered by assuming that
rates do not depend on the previous transition ($\kappa_{\ NM}^{PN}
\rightarrow \kappa^{P}_{\ N}$), and then summing over $M$ ($f^{N} \equiv
\sum_{M} f^{NM}$) to yield
\be
\frac{df^{N}}{dt} = \sum_{P} - \kappa^{P}_{\ N} f^{N} + \sum_{M} \kappa^{N}_{\ M} f^{M}. 
\ee
Note that there are $n_{v}$ equations for $f^{N}$, but $n_{v} (n_{v} - 1)$ equations for  $f^{NM}$, reproducing the fact that there are $(n_{v}-1)$ more degrees of freedom in a transition-based framework.

The procedure given by~\cite{Garriga:2005av} for counting the total number of bubbles of type $N$ nucleated in a background $M$ before some ($NM$-dependent) time cutoff can be generalized straightforwardly: the number of such bubbles formed per unit time would be given by the formation rate $\sum_{L} \kappa_{\ ML}^{NM} f^{ML}$ of comoving volume fraction $f^{NM}$, divided by the asymptotic comoving volume of bubbles of $N$.  $f^{NM}$ itself can be calculated by formulating Eq.~\ref{eq-rates} as a matrix problem, in a manner similar to that for the standard rate equations presented in~\cite{Garriga:2005av}. 

Bubble-counting measures may be extended to longer histories (of bubbles within bubbles within bubbles...) in a similar manner as for transitions along a worldline.

\section{Conclusions}
\label{sec-conclusions}

Many cosmological observables depend upon how the inflaton evolves to the minimum of its potential, which in turn depends on how that minimum's basin of attraction was entered. We have therefore argued that a measure for eternal inflation should assign weights to {\em transitions} between vacua, as opposed to existing measures that count vacua regardless of how they were reached. Moreover, a measure on transitions is a more natural way to apply many of the ``anthropic'' conditionalizations being considered today (baryons, galaxies, entropy produced, etc.), since these also generally depend upon the transition type rather than simply the vacuum considered.

We showed how two proposed measures  -- counting either vacuum entries by a worldline, or the total number of bubbles of different vacua in an eternally inflating spacetime -- could be modified to count transitions as opposed to vacua, as well as how the transition formalism could be extended to allow for history-dependent transition rates, and to provide probabilities for longer histories.

\begin{acknowledgments}

We thank A. Lewis for helpful discussions. AA, SG, and MJ were supported, respectively, by  a ``Foundational Questions in Physics and Cosmology"
grant from the John Templeton Foundation, by PPARC, and by the ARCS Foundation.

\end{acknowledgments}

\bibliography{histories.bib}

\end{document}